\documentclass[12pt]{iopart}
\usepackage{iopams}
\usepackage{epsf}
\usepackage{euscript}
\usepackage{graphicx}

\renewcommand{\epsilon}{\varepsilon}
\newcommand{\tg}{{\rm tg}}

\begin{document}
\jl{1}

\title[Can One Hear the Shape of a  Graph?]%
{Can One Hear the Shape of a  Graph?}
\author{Boris Gutkin\ and Uzy Smilansky}
\address{ Department of Physics of Complex Systems,
The Weizmann Institute of Science, 76100 Rehovot, Israel}
\address{\vspace*{\baselineskip} E-mail: \texttt{Uzy.Smilansky@weizmann.ac.il}}
\begin{abstract}
We show that the spectrum of the  Schr\"odinger operator on a finite,
metric  graph
determines {\it uniquely} the connectivity matrix and  the bond lengths,
provided that the
lengths are non-commensurate and the connectivity is simple (no parallel
bonds between
vertices and no loops connecting a vertex to itself).  That is, one can
hear the shape of
the graph! We also consider  a related inversion problem: A compact graph
can be converted
into a scattering system by attaching to its vertices leads to infinity. We
show that the
scattering phase determines uniquely the compact part of the graph, under
similar
conditions as above.
\end{abstract}
\maketitle

\section{Background and notations}

 The question ``Can one hear the shape of a drum?",  was posed by  Marc Kac
\cite {Kac66} as
a paradigm example for a class of problems which is of fundamental
importance in many
physical applications: Given the quantum spectrum, can one deduce uniquely
the basic
interactions or the geometric constraints  which specify the system? In
Kac's original
paper, this inversion problem is formulated for Laplacians on compact
domains with boundary
conditions (billiards), and in spite of approximately fifty years of active
research, the
complete answer is still an enigma. If the billiard boundary is allowed to
have corners,
one can draw different yet isospectral billiards \cite {corners92,Chap95}.
Isospectral microwave cavities were also constructed \cite{sund94} using
the same models.
Recently, it was  shown  that Kac's question is answered in the affirmative for
simply connected doamins with analytic boundaries with  some symmetry
restrictions \cite
{Zel00} . For boundaries in intermediate classes of smoothness the answer
is not known.
The  existence of isospectral systems was investigated for  Laplacians on
closed Riemannian
manifolds \cite {sunda85} and for discrete Laplacians which are formed by
the connectivity
matrices of graphs \cite {brooks99}. In both cases, elaborate techniques
were devised to
identify large sets of different, but isospectral systems. However, if the
domains are
analytic surfaces of revolution, the spectrum determines the manifold uniquely
\cite {Zel98}.

In the present note we address yet another class of problems which go under
the name ``quantum graphs". The Schr\"odinger operator consists of the
one-dimensional Laplacian on the bonds, endowed with boundary conditions on the
vertices, which ensure that the spectra are real and discrete. (see
\cite{KS99} for
a detailed discussion and list of references). These systems recently
attracted the
attention of the quantum chaos community, since, in spite of their structural
simplicity, their spectra display the complexity and statistics which
characterize
generic quantum chaotic systems such as billiards. Moreover, an exact trace
formula
\cite {Roth83,KS97} can be written down in terms of the periodic orbits (PO) on
the graph, which is analogous to the semiclassical trace formula for chaotic
systems. So far, this exact trace formula paved the way to the  study of
spectral
statistics by applying combinatorial methods to periodic orbit expansions
\cite{SS00,BK00}. Here we shall show that it also provides the key to the
affirmative answer to Kac's question, when applied to graphs.

\medskip

 A graph ${\cal G}$ consist of $V$ {\it vertices} connected by $B$ {\it
bonds}.  The
$V\times V$ {\it connectivity matrix} is defined by:

\begin{equation}
     C_{i,j}=  {\rm number\ of\ bonds\  connecting\  the\   vertices}\  i\
{\rm
and}\   j\  .
\end{equation}

\noindent
 A graph is {\it
simple} when for all $i,\ j \ :\  C_{i,j}\in [0,1]$ (no parallel
connections) and
$C_{i,i}=0$ (no loops). The {\it valence} of a vertex is
$v_i =
\sum_{j=1} ^V C_{i,j}$ and the  number of bonds is $B={1\over 2}
\sum_{i,j=1} ^V C_{i,j} $.  We denote the bonds connecting the vertices $i$
and $j$
by $b=[i,j]$.  The notation $[i,j]$ and the letter $b$ will be used whenever
we refer to a bond without specifying a {\it direction}. Hence,
$b=[i,j]=[j,i]$. To
any vertex
$i$ we can assign in a unique way the set $s^{(i)}$ of bonds which emanate
from it:
\begin{equation}
s^{(i)} =\{{\rm all \
bonds} \ [i,k]   :  C_{i,k}=1 \}\  ;\ card[s^{(i)}] = v_i .
\label{eq:star}
\end{equation}
 We shall refer to   $s^{(i)}$ as a  {\it
topological star}, since its definition does not require any metric
information.
For simple graphs,
\begin{equation}
C_{i,j} = C_{j,i} =card[ s^{(i)}\cap s^{(j)}] \ .
\label{eq:connstar}
\end{equation}
 {\it Directed bonds} will be
denoted by $d=(i,j)$, and we  use the convention that the bond is
directed from the first to the second index. The round brackets, and the letter
$d$ are reserved to denote the directed bonds. The reverse bond to
$d=(i,j)$ is denoted by $\hat d =(j,i)$. In analogy to  (\ref {eq:star}), we
define the sets of directed bonds {\it outgoing} from or {\it incoming} to $i$:
\begin{eqnarray}
&&s^{(i,+)} =\{{\rm all \ directed \
bonds} \ (i,k)   :  C_{i,k}=1 \} \nonumber  \\
&&s^{(i,-)} =\{{\rm all \ directed \
bonds} \ (k,i)   :  C_{k,i}=1 \}   \ ,
\label{eq:star+-}
\end{eqnarray}
respectively. Note: $
card[s^{(i,+)}] =  card[s^{(i,-)}] =v_i$. A directed bond $d'$ is defined to
{\it follow} the directed bond $d$ at the vertex $i$  if $d\in s^{(i,-)}$
and $d'\in s^{(i,+)} $. It is convenient to define a directed connectivity matrix $F$,
with
$F_{d,d'} = 1$ if $d$ follows $d'$, and $ F_{d,d'} = 0$ otherwise.
Any directed bond $d=(i,j)$ belongs to the two sets, $ s^{(i,+)}$ and
 $s^{(j,-)}$,  and  $C_{i,j} = card[s^{(i,+)}\cap s^{(j,-)}] =  card[s^{(j,+)}\cap s^{(i,-)}]$.

 We shall now define the Schr\"odinger operator on ${\cal G}$, and collect
a few
facts which will be used in the sequel. For a detailed exposition see e.g.,
\cite
{KS99}.  We assign  the natural metric to the bonds, and each bond is
endowed with a
length $L_b$. The Schr\"odinger operator consists of
\begin{equation}
\left [-{\rm i}{ {\rm d\ }\over {\rm d}x} - A_b\right ]^2
\label{eq:schroedinger}
\end{equation}
on all the bonds, and $A_b$ are constants which are introduced to
break the symmetry between the counter propagating waves on the bond. The
differential operator is supplemented by boundary conditions which are
imposed in
the following way. The most general solution of the Schr\"odinger equation
on any
bond
$b=[i,j]$  is
\begin{equation}
\psi_b(x) = a_{\hat d} {\rm e}^{ i(k+A_b)x} + a_d {\rm e}^{-i(k-A_b)x} \ \
; \ \
\forall\ b\in s^{(i)} \ ,
\label{eq:waves}
\end{equation}
where $d$, and $\hat d$ denote the two directions on the bond $b$, and
$a_d,\ a_{\hat
d} $ are yet unspecified complex numbers.  The boundary conditions impose the
relations
\begin{equation}
\forall d\in  s^{(i+)}\ :  \ \ a_d = \sum_{d'\in s^{(i-)}}\sigma^{(i)}_{d,d'}\
a_{d'}
\ ,
\end{equation}
where the {\it vertex scattering matrix} $\sigma^{(i)}$ is a  unitary,
 $v_i\times v_i$ matrix. $\sigma^{(i)}$ can be either an arbitrary  $k$
independent
unitary matrix, or of the general class of (possibly $k$ dependent)
matrices derived from
matching conditions at the vertices \cite {KS99,KoSch99}. The matrix
$\sigma^{(i)}$ will be called  {\it properly connecting} if none of its
matrix elements
vanishes. The {\it transition probabilities}
 $
W^{(i)}_{d,d'} = |\sigma^{(i)}_{d,d'}|^2
$
appear in the classical dynamics analogue, which is a random walk model on the
directed  bonds, with transition probablilities $
W^{(i)}_{d,d'} $. As an example,
consider the  most commonly used and discussed quantum graphs, where
 the vertex scattering matrices are derived by imposing Neumann boundary
conditions at the vertices. In this case
$
\sigma^{(i)}_{d,d'} = -\delta_{d',{\hat d}} +{2\over v_i} \ ,
$
and but for the trivial vertices with $v_i=2$, these scattering matrices
are properly
connecting.

Combining the boundary conditions for all the vertices, results in a
secular equation
for the wavenumbers  $k_n$ which consists of the  spectrum
 of the Schr\"odinger operator. The secular equation reads,
\begin{equation}
\det(I-S(k))=0
\label{secular}
\end{equation}
where  $S(k)$ is a $2B\times 2B$ unitary matrix, whose rows and
columns are labeled by the directed bond labels.  It is defined as
\begin{equation}
{S}_{d,d'}(k) = F_{d,d'}
{\rm e}^{i(k+A_d) L_{d} }\sigma^{(i)}_{d,d'}\ ,
\label{Sdef}
\end{equation}
The index $i$ denotes the vertex at which $d$ follows $d'$, and
$L_d=L_{\hat d}$,
is the bond length which is independent of the direction of propagation.
However
$A_d=-A_{\hat d}$, which distinguishes between the directions of propagation.
$S (k)$ can be interpreted as a quantum  evolution operator describing
the scattering of waves with wave number $k$ between connected bonds. The wave
which scatters at the vertex $i$ from $d'\in s^{i,-}$ to $d\in s^{i,+}$ with an
amplitude  $\sigma^{(i)}_{d,d'}$, gains the phase $(k+A_d) L_{d} $ during the
propagation along the outgoing bond $d$. With the next application of
$S(k)$ the
wave scatters again on the vertex to which $d$ is directed. If all
$\sigma^{(i)}$
matrices are properly connecting, the waves on ${\cal G}$ propagate between
all the
topologically connected bonds. The unitarity of the matrices
$\sigma^{(i)}$ guarantees that $S(k)$ is unitary for real $k$. This is also
why the
Schr\"odinger operator is self-adjoint and its spectrum is pure-point and
unbounded on the real line.

 The secular equation (\ref {secular}), with the form  (\ref {Sdef}) for $S(k)$
leads naturally to the exact trace formula \cite {Roth83,KS97}
\begin{eqnarray}
 d(k)  &\equiv&   \sum_{n} \delta (k-k_n) \nonumber \\
 &=& {\frac {{\cal L}}{\pi }}+
 {\frac {1}{2\pi} }\sum_{p}({\cal A}_{p} {\rm e}^ {ikl_p} +{\cal
A}^{\ast}_{p} {\rm
e}^ {-ikl_p} )
\label {trace}
\end{eqnarray}
 where ${\cal L}= \sum_{b=1}^B L_b$ is the total length of ${\cal G}$. The sum
extends over all the  PO's on the graph, and goes
over primitive PO's as well as  their repetitions.  The length of a  PO is
denoted by
$l_p$.  The coefficient ${\cal A}_p$ is  a product over all the scattering
amplitudes
$\sigma^{(i)}_{d,d'}$, encountered along the PO, times the length of the
primitive PO of which $p$ is a repetition. It is also endowed with a phase
factor
$\exp(i\sum A_d L_d)$. The length of a  PO is
\begin{equation}
 l_p = \sum_{b=1}^B\ q^{(p)}_b\ L_b \ \ \ , \ \     q_b\in N_0
\label{eq:length}
\end{equation}

The {\it length spectrum} is the Fourier transform of (\ref {trace}),
\begin{eqnarray}
d(l)& \equiv& \sum_{n} {\rm e}^{-ilk_n} =
  2{\cal L} \delta(l)
 +\sum_{p}\left ({\cal A}_p
   \delta \left( l- l_p \right )+{\cal A}^{\ast}_p
   \delta \left( l+ l_p \right )\right ).
 \label{eq:lspectrum}
\end{eqnarray}

\section{Hearing the shape of the  graph}

After reviewing the necessary background, we are in position to formulate
the answer to
Kac's question for graphs.

\noindent {\it Theorem 1:}  The spectrum of a Schr\"odinger operator on a
metric
graph determines uniquely the graph connectivity and the length of the
bonds, provided that

\noindent - the graph is finite and simple,

\noindent - the bond lengths are rationally independent,

\noindent - the vertex scattering matrices are properly connecting.

Before presenting the formal proof, we shall sketch its  main idea:  The length
spectrum(\ref {eq:lspectrum}) is constructed from the Schr\"odinger
spectrum. Its
singularity at $l=0$ provides the total length of the graph ${\cal L}$,
which, in
turn, gives an upper bound to the individual bond lengths. The conditions
that the
lengths are rationally independent, and that whenever $d$ follows $d'$ at
vertex
$i$, $\sigma^{(i)}_{d,d'} \ne 0 $, ensure that  the lengths of all the PO's
 which are consistent with the graph connectivity,  appear as $\delta $
singularities in the length spectrum, and their corresponding coefficients
${\cal
A}_p$ do  not vanish. Moreover, PO's $p,p'$ which traverse the bonds different
numbers of times ( $q^{(p)}_b \ne q^{(p')}_b$ for some $b$ in (\ref
{eq:length}))
have distinct lengths. We use these facts to isolate two classes of periodic
orbits. The  $2$-PO's of the type $[d,{\hat d}]$ provide the lengths of the
bonds.
The $4$-PO's of the  type
$[d^{(l)},{\hat d^{(l)}}, d^{(m)},{\hat d^{(m)}}]$ for a fixed $l$ and for all
possible $m$   are used to identify the topological star at the vertex $i$
where
$d^{(m)}$ follows ${\hat d^{(l)}}$. Once all the topological stars
$s^{(i)}$  are found,  we  apply  (\ref{eq:connstar}) to get the
connectivity.  Thus, the shortest, and structurally simplest PO's
are used to extract the parameters which determine the ``shape" of ${\cal
G}$.

The  proof of the theorem  proceeds as follows:

\noindent {\it i)}. Use the spectrum to construct the sum
(\ref{eq:lspectrum}). Its
singular support $R$ consists of $l=0$ and the infinitely many points $l_p$
of the form
(\ref {eq:length}) on the real axis. The weight of the singularity at $l=0$
is $2{\cal
L}$.

\noindent  {\it ii)} Generate a finite set $P$ which consists of all the
strictly positive lengths $l_p\in R$, which are less than $ 2{\cal L}$.
Exclude
any length which can be expressed as an integer multiples of any other
length, with
a multiplier larger than 1. $P$ is the (finite) set of lengths of primitive
PO's on ${\cal G}$, with lengths in $(0,{2\cal L}]$.

\noindent {\it iii)} Define the {\it basis} $Q  \subset  P $, which
consists of the
minimum number of lengths $\lambda_q \in P$ such that
\begin{equation}
\forall l_p\in P\  , \  l_p = {1\over 2} \sum_{q=1}^{card[Q] } n_q\
\lambda_q \ , \ \
n_q\in N_0
\label{eq:qdef}
\end{equation}
$P$ is a finite set, and therefore $Q$ can be constructed in a finite
number of steps.
Because of the rational independence of the bond lengths, the basis is
composed of the
lengths of $2$-PO's $[d,\hat d]$, with
$\lambda_q= 2 L_q$ where
$L_q$ are the lengths of the bonds. $card[Q] =B$ is the number of bonds,
and $\sum \lambda _q= 2{\cal L}$.

\noindent{\it iv)} In what follows we shall construct {\it metric stars}
which will be shown
to be in one-to-one  correspondence with the topological stars
(\ref{eq:star}):
Define two lengths  $\lambda_q$ and $\lambda_{q'}
\in Q$ to be {\it connected} if $\lambda_q \ +\ \lambda_{q'} \in P$. A
length in $Q$ cannot
be connected to itself since repetitions where excluded from $P$. The
lengths in $Q$
which are not connected to any other length, correspond  to  disconnected
bonds, which
are identified this way, and excluded from the  susequent analysis.  Given
a length
$\lambda_r\in Q$, consider all the {\it clusters}
$C_m(\lambda_r) \subset  Q$ satisfying the following requirements: (a)
 $\lambda_r \in C_m(\lambda_r)$. (b) All the lengths in   $C_m(\lambda_r)$ are
pairwise connected. (c)  $C_m(\lambda_r)$ is maximal -  adding another
$\lambda_s \in Q$,
will violate (b).  Denote the number of $C_m(\lambda_r)$ by
$N(\lambda_r)$. Metric stars are special clusters defined as follows. If
for some
$\lambda_s$, $N(\lambda_s)=1$  we shall define $ \{\lambda_s \}$, as a {\it
metric star}
consisting  of a single element. If $card[C_m(\lambda_s) ]=2$ then it is a
metric star with
two elements. All other $C_m(\lambda_r)$ are metric stars if for any three
elements $
\lambda_l,\  \lambda_{m},\ \lambda_{n}  \in C_m(\lambda_r),  \
\lambda_l+\lambda_{m}
+\lambda_{n} \in P$.
$P$ is pruned of repetitions. Hence, this condition excludes  repeated
triangular PO's
from being considered as stars. Using all the $\lambda_r\in Q$ as reference
lengths,  we
construct all the different  metric stars in
$Q$. The bond lengths  are rationally independent, thus, the metric stars
$S^{(i)} $are in
$1\leftrightarrow 1$ correspondence with  the topological stars
(\ref{eq:star}). The
proof is completed  by writing the connectivity matrix in
 analogy with (\ref{eq:connstar}):
\begin{equation}
C_{i,j} =card[ S^{(i)}\cap S^{(j)}] \ ;  \ L_{[i,j]}= {1\over 2}\left(
S^{(i)}\cap S^{(j)}
\right ) \ .
\label{eq:connmetstar}
\end{equation}
 The length spectrum can be used to get  more information about the graph
\cite
{car99}.    The coefficients of the $\delta$ singularities provide
constraints on the
values of the elements of the vertex scattering matrices. Assuming that the
$\sigma^{(i)}$ are symmetric, one can determine the transition
probabilities $W^{(i)}_{d,d'}$ from the coefficients of the
lengths of three  classes of simple PO's. Denote
$b=[i,j], b'=[i,k], b''=[i,l]$ and correspondingly $ d=(i,j), d'=(i,k),
d''= (i,l)$. They
belong to the same star.  The three  types  of PO's are
$[d,\hat d]$, $[d',{\hat d'},d,{\hat d}]$, and $[d'',{\hat d''},d',{\hat
d'},d,{\hat d}]$,
with  lengths $2L_b$, $2(L_b+L_{b'})$, and $2(L_b+L_{b'}+L_{b''})$,  with
corresponding
coefficients
 \begin {eqnarray}
& &{\cal A} _{[d,{\hat d}]} = 2\ L_b\ \sigma^{(i)}_{d,{\hat
d}}\sigma^{(j)}_{{\hat d},d}
\label {eq:type1} \\
& & {\cal A}_{[d'{\hat d'},d,{\hat d}]} = 2\ (L_b+L_{b'})\
\sigma^{(i)}_{{d',\hat d}}
\sigma^{(k)}_{{\hat d},d}
\sigma^{(i)}_{d, {\hat d'}}
\sigma^{(j)}_{{\hat d'},d'}
\label {eq:type2}\\
& & {\cal A}_{[d'',{\hat d''},d',{\hat d'},d,{\hat d}]} =  4 \
(L_b+L_{b'}+L_{b''})\
\times  \label {eq:type3}
\\
&&\ \ \ \ \ \ \ \ \ \ \ \ \ \ \ \ \ \  \ \ \ \ \   \sigma^{(i)}_{{d'',\hat d}}
\sigma^{(k)}_{{\hat d},d}
\sigma^{(i)}_{d,{\hat d'}}
\sigma^{(j)}_{{\hat d'},d'}
\sigma^{(i)}_{d',{\hat d''}}
\sigma^{(l)}_{{\hat d''},d''}   \nonumber
\end{eqnarray}
Dividing (\ref {eq:type2}) and (\ref {eq:type3}) by (\ref {eq:type1})  we
get ratios of
the matrix elements of a single $\sigma^{(i)}$ of the form
\begin {equation}
{{\cal A}_{[d'{\hat d'},d,{\hat d}]}\over {\cal A}_{[d,{\hat d}]}
{\cal A}_{[d',{\hat d'}]}} {2L_b L_{b'}\over (L_b+L_b')}  =
{\sigma^{(i)}_{{d',\hat d}}
\sigma^{(i)}_{d, {\hat d'}}\over \sigma^{(i)}_{{\hat d},d} \sigma^{(i)}_{{\hat
d'},d'}} \ .
\label {eq:coef3}
\end{equation}
Ratios of  products of three matrix elements can also be formed. Using
this information one can construct the matrices $\sigma^{(i)}$ uniquely, up
to a phase
factor since the ratios  of the type (\ref {eq:coef3}) are invariant under
a right and left
multiplication by arbitrary diagonal matrices.  Hence one obtains only
absolute values, in other words, the transition probabilities
$W^{(i)}_{d,d'}$.
When $\sigma^{(i)}$ are real,  $L_b A_b {\rm mod} 2\pi$ can also be
computed from the weights of simple PO's.

\bigskip
\section {Application to non-compact (scattering) graphs}
A related inversion problem, is encountered when the
compact graph is turned into a scattering system by coupling to its
vertices leads
(bonds) to infinity \cite {KS00}. It is assumed throughout that each vertex
can be attached to {\it at most} one lead. The definition of the
Schr\"odinger operator
can be extended naturally to the non-compact case, and the coupling to the
leads is
achieved by modifying the vertex scattering matrices to include coupling to the
leads. Let $N$ be the number of vertices to which leads are attached, and
the leads
are denoted by the index of the vertices to which they are connected. Then, for
every wavenumber
$k$ one can find a solution to the Schr\"odinger equation which on the
leads takes
the form
\begin{equation}
\psi_l(x) = I_l (k){\rm e}^{-ikx} + O_l(k) {\rm e}^{+ikx}
\label{eq:leadswf}
\end{equation}
and the outgoing amplitudes $O_l(k)$ are related to the incoming amplitudes
$I_l
(k)$ by
\begin{equation}
O_l(k) = \sum_{l'=1}^N T_{l,l'}(k) I_{l'}(k)
\label{eq:Tmatrix}
\end{equation}
where $T(k)$ is the $N\times N$ scattering matrix. The conservation of flux
ensures
the unitarity of $T(k)$. An explicit expression for the scattering matrix is
given in \cite {KS00}. Here, we shall be concerned with the {\it scattering
phase}
defined by
\begin{equation}
\Theta(k) = {1\over2\pi  i} \log \left [ \det T(k)\right]
\label{eq:scatphase}
\end{equation}
 The scattering phase is the counterpart of the spectral counting function for
compact systems. This can be best seen by considering the function
$d_R(k)={{\rm
d}\Theta(k) \over {\rm d}k}$. It assigns a normalized Lorentzian to each
scattering resonances, and it is the counterpart of the spectral density
(\ref {trace}). An exact trace formula for $d_R(k)$ was derived in \cite
{KS00}. It
reads,
\begin{eqnarray}
 d_R(k)  &\equiv&   \sum_{n} {1\over \pi} {\gamma_n\over (k-k_n)^2+\gamma^2_n}
\nonumber
\\
 &=& {\frac {{\cal L}}{\pi }}+
 {\frac {1}{2\pi} }\sum_{p}(\tilde {\cal A}_{p} {\rm e}^ {ikl_p} +\tilde {\cal
A}^{\ast}_{p} {\rm e}^ {-ikl_p} )
\label {tracer}
\end{eqnarray}
where $k_n-{\rm i}\gamma_n$ are the poles (resonances) of $T(k) $ in the
complex $k$ plane. As in the compact case,  ${\cal L}=
\sum_{b=1}^B L_b$ is the total length of the bonds in the compact part of
the graph.
 The sum  extends over all the  PO's (periodic orbits) on the compact part of
the graph, and goes over primitive PO's as well as  their repetitions.  The
length of a  PO is denoted by $l_p$.  The amplitude $\tilde {\cal A}_p$ is  a
product over all the scattering amplitudes
$\tilde \sigma^{(i)}_{d,d'}$, encountered along the PO, times the length of the
primitive PO of which $p$ is a repetition. It is also endowed with a phase
factor
$\sum A_d L_d$. The vertex scattering matrices $\tilde \sigma^{(i)}$ in the
space
of the directed compact bonds are sub-unitary ($|\det \tilde \sigma^{(i)}|
\le 1$).
To complete $ \tilde \sigma^{(i)}$ to a unitary matrix
 a row and a colomn should be added, to take into account the coupling of the
compact bonds to the lead. For this reason, the rows of the classical
transition matrix
$\tilde W^{(i)}_{d,d'} = |\tilde \sigma^{(i)}_{d,d'}|^2$ do not sum up to
unity, which take
into account the flux which escapes through the leads.   The same mechanism is
 responssible also to the  reduction of
$|\hat {\cal A}_p| $ relative to the corresponding
$|{\cal A}_p|$. This accelerates the convergence of (\ref{tracer}) and
pushes its poles
away from the real axis. A length spectrum similar to (\ref {eq:lspectrum})
is obtained
from the Fourier transform of (\ref {tracer}), and its singular support
consists of
$0$ and the length  spectrum of the compact part of the graph. The
following result
can be proved using the same argument as above:

\noindent{\it Theorem 2:} The  scattering phase $\Theta(k)$ of a non-compact
metric graph,  determines uniquely the connectivity and the bond lengths in
the compact
part of the graph, provided that

\noindent - the graph is composed of a finite and simple compact part, with
leads to
infinity. A vertex can be coupled to at most one lead.

\noindent - the bond lengths in the compact part are rationally

\noindent $\ $ independent,

\noindent - the vertex scattering matrices are properly connecting.

The proof follows  {\it verbatim} the proof of {\it theorem 1}.

%
%
\begin{figure}
  \begin{center}
    \includegraphics[scale=.6] {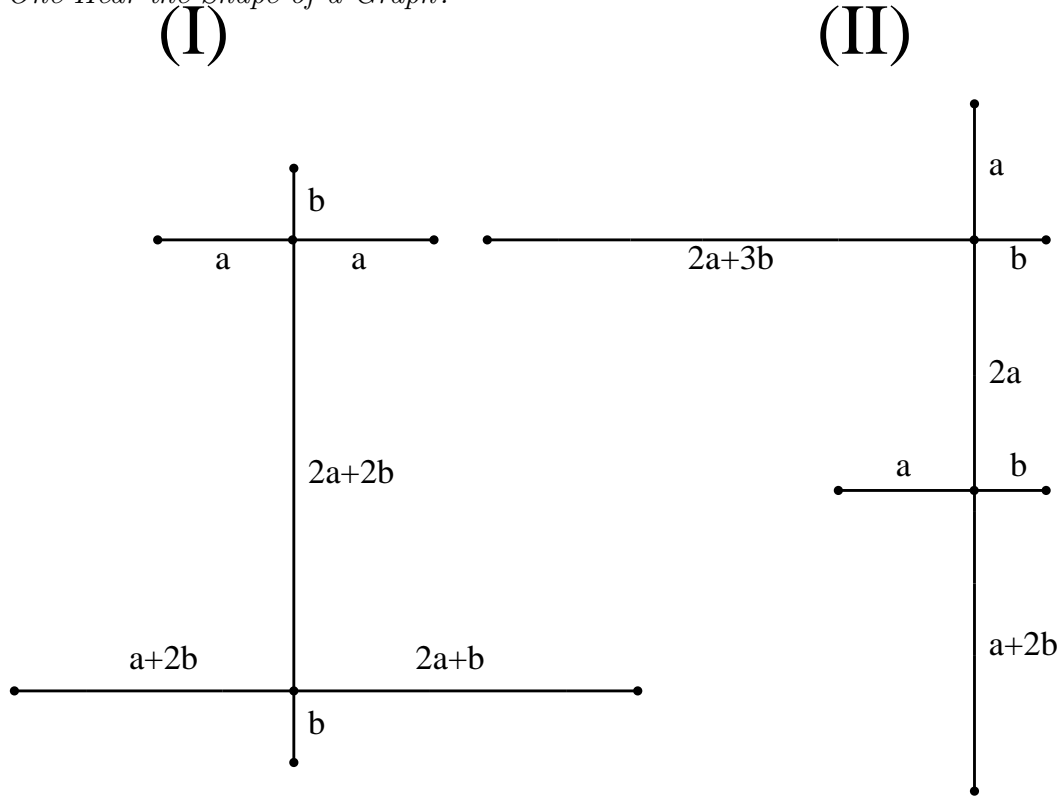}
  \end{center}
  \caption{Two isospectral graphs. The bond lengths are  expressed in
    terms of the two arbitrary lengths $a$ and $b$.}
  \label{fig:1}
\end{figure}
%

\section{An example of isospectral graphs}

We have shown that the spectral density (for compact graphs) and the
resonnance density (for  scattering graphs) determine uniquely the
bond lengths and the the connectivity,  provided that the bond lengths are
rationally independent, and  the vertex scattering matrices are properly
connecting.
We shall now construct a non trivial example of two different, yet
isospectral graphs,
obtained when the requirement of rational independence is relaxed.

Extending the geometric construction of \cite {{Chap95}} we produced the
pair of graphs
shown in Figure 1. Their bond lengths are rational combinations of two
arbitrary lengths $a$
and $b$, as shown in the figure. Neumann boundary conditions are imposed at
the vertices.
The spectra of the two graphs are the zeros of the corresponding  secular
functions (\ref
{secular}), which for the present cases read:
\begin{eqnarray}
 &&Z_{(I)}(k)= \tg (2\,\left( {a} + {b} \right) \,k)  +   \\
&&
  {\frac{2\,\tg ({a}\,k) + 2\,\tg ({b}\,k) +
      \tg (\left( 2\,{a} + {b} \right) \,k) +
      \tg (\left( {a} + 2\,{b} \right) \,k)}{1 -
      \left( 2\,\tg ({a}\,k) + \tg ({b}\,k) \right) \,
       \left( \tg ({b}\,k) +
         \tg (\left( 2\,{a} + {b} \right) \,k) +
         \tg (\left( {a} + 2\,{b} \right) \,k) \right) }}
\nonumber \\
\nonumber \\
&& Z_{(II)}(k)=
\tg (2\,{a}\,k)   \\
&& {\frac{2\,\tg ({a}\,k) +
      2\,\tg ({b}\,k) + \tg (\left( {a} + 2\,{b} \right)
          \,k) + \tg (\left( 2\,{a} + 3\,{b} \right) \,k)}{1 -
      \left( \tg ({a}\,k) + \tg ({b}\,k) +
         \tg (\left( {a} + 2\,{b} \right) \,k) \right) \,
       \left( \tg ({a}\,k) + \tg ({b}\,k) +
         \tg (\left( 2\,{a} + 3\,{b} \right) \,k) \right) }}
\nonumber
\end{eqnarray}

It is not an easy matter to show directly that these two different
functions vanish at
exactly the same values of $k$. However, the theorem which underlies their
construction
\cite {Chap95} guarantees this fact.

\section{Acknowledgements}

Support by the Minerva Center for Nonlinear Physics, the  Israel
Science Foundation and the Minerva Foundation are acknowledged. We thank
U.~Gavish K.
Naimark and M. Solomiak for valuable comments.

\end{document}